\documentclass{lmcs}
\pdfoutput=1

\usepackage{lastpage}
\lmcsdoi{20}{3}{15}
\lmcsheading{}{\pageref{LastPage}}{}{}%
{May~27,~2022}{Aug.~12,~2024}{}

\usepackage[utf8]{inputenc}


\usepackage{amssymb}
\usepackage{tikz}

\newcommand{\tuple}[1]{\langle #1 \rangle}
\newcommand{\emptyword}[0]{\varepsilon}
\newcommand{\trans}[1]{\mathchoice{\xrightarrow{#1}}{\xrightarrow{\smash{\lower1pt\hbox{$\scriptstyle #1$}}}}{\xrightarrow{#1}}{\xrightarrow{#1}}}
\newcommand{\tranr}[1]{\mathchoice{\xleftarrow{#1}}{\xleftarrow{\smash{\lower1pt\hbox{$\scriptstyle #1$}}}}{\xleftarrow{#1}}{\xleftarrow{#1}}}
\newcommand{\extraone}[0]{\mathord{\lower-0.25pt\hbox{$\vartriangle$}}}
\newcommand{\extratwo}[0]{\mathord{\hbox{\scalebox{0.75}{$\square$}}}}

\DeclareUnicodeCharacter{3C1}{\rho}                     
\DeclareUnicodeCharacter{2113}{\ell}                    
\DeclareUnicodeCharacter{2026}{\ldots}                  
\DeclareUnicodeCharacter{2115}{\mathbb{N}}              
\DeclareUnicodeCharacter{2190}{\leftarrow}              
\DeclareUnicodeCharacter{2192}{\rightarrow}             
\DeclareUnicodeCharacter{2208}{\in}                     
\DeclareUnicodeCharacter{221E}{\infty}                  
\DeclareUnicodeCharacter{222A}{\cup}                    
\DeclareUnicodeCharacter{223C}{\sim}                    
\DeclareUnicodeCharacter{2260}{\neq}                    
\DeclareUnicodeCharacter{228E}{\uplus}                  
\DeclareUnicodeCharacter{22A5}{\bot}                    
\DeclareUnicodeCharacter{22EF}{\cdots}                  
\DeclareUnicodeCharacter{230A}{\lfloor}                 
\DeclareUnicodeCharacter{230B}{\rfloor}                 
\DeclareUnicodeCharacter{2A7E}{\geqslant}               
\DeclareUnicodeCharacter{1D4AF}{\mathcal{T}}            

\begin{document}

\title[Pushdown compression and normality]{Deterministic pushdown automata\texorpdfstring{\\}{} can compress some normal sequences}

\author[O.~Carton]{Olivier Carton\lmcsorcid{0000-0002-2728-6534}}[a,b]
\author[S.~Perifel]{Sylvain Perifel\lmcsorcid{0000-0002-5782-820X}}[a]

\address{Université Paris Cité, CNRS, IRIF, F-75013, Paris, France}
\email{Olivier.Carton@irif.fr, Sylvain.Perifel@irif.fr}

\address{Institut Universitaire de France}

\begin{abstract}
  In this paper, we give a deterministic one-to-one pushdown transducer and
  a normal sequence of digits compressed by it.  This solves positively a
  question left open in a previous paper by V. Becher, P. A. Heiber and the
  first author.
\end{abstract}

\maketitle

\section{Introduction}

A real number is normal to an integer base if, in its infinite expansion in
that base, all blocks of digits of the same length have the same limiting
frequency.  Émile Borel~\cite{Borel09} defined normality more than one
hundred years ago to formalize the most basic form of randomness for real
numbers.  Many of his questions are still open, such as whether any of
$\pi, e$ or~$\sqrt{2}$ is normal in some base, as well as his conjecture
that the irrational algebraic numbers are normal to each
base~\cite{Borel50}.  This motivates the search for new characterizations
of the concept of normality.

One characterization is based on finite state machines. A sequence of
digits is normal if and only if it cannot be compressed by lossless finite
transducers (also known as finite-state compressors).  These are
deterministic finite automata augmented with an output tape with injective
input-output behavior.  The compression ratio of an infinite run of a
transducer is defined as the $\liminf$, over all its finite prefixes, of
the ratio between the number of symbols written and the number of symbols
read so far.  A given sequence is said to be compressed by a given
transducer if the compression ratio it achieves is less than~$1$.

A direct proof of the incompressibility characterization of normal
sequences can be found in~\cite{BecherHeiber13}.  However, the result was
already known, although by indirect and more involved arguments. For
instance, combining results of Schnorr and Stimm~\cite{Schnorr71} and Dai,
Lathrop, Lutz and Mayordomo~\cite{Dai04} yields an earlier proof: the
characterization of normality given in~\cite{Schnorr71} is based on
martingales and the equivalence between martingales and compressibility is
shown in~\cite{Dai04}.  It is also proved
in~\cite{DotyMoser06,SheinwaldLempelZiv95} that compression ratio and
decompression ratio coincide.

The notion of incompressibility by finite state machines is quite robust:
adding some feature to one-to-one transducers does not allow them to compress
normal sequences. It is proved in \cite{BecherCartonHeiber15} that
non-deterministic non-real-time transducers, with no extra memory or just a
single counter, cannot compress any normal sequence. Non-real-time means here
that the value of the counter can be incremented and decremented without
consuming any input symbol. It is also shown in \cite{CartonHeiber15} that
two-way transducers cannot compress normal sequences. Adding too much memory
yields compressibility results: it is clear that Turing complete machines can
compress computable normal sequences like the Champernowne
sequence~\cite{Champernowne33}. This includes non-real-time transducers with
at least two counters. Note however that Turing completeness is not necessary.
For instance, it is shown in~\cite{LathropStrauss97} that some normal
sequences are compressed by Lempel-Ziv algorithm. Combining non-determinism
with a single stack also yields compressibility of some normal sequence.
Results given in~\cite{BecherCartonHeiber15} are summarized in
Table~\ref{tbl:compress}. One question left open was whether a deterministic
pushdown transducer can compress a normal sequence, that is the question mark
in Table~\ref{tbl:compress}, where (T) means Turing-complete. In this paper,
we answer this question positively.

\begin{table} \label{tbl:compress}
  \begin{center}
  \begin{tabular}{|l|c|c|} \hline
  Finite-state transducer   & det. &  non-det. \\ \hline
  No extra memory           & N & N  \\ \hline
  One counter               & N & N  \\ \hline
  One stack                 & \textbf{?} & Y  \\ \hline
  More than one counter     & Y (T) & Y (T)  \\ \hline
  One stack and one counter & Y (T) & Y (T) \\ \hline
  \end{tabular}
  \end{center}
  \caption{Compressibility by different kinds of transducers.}
\end{table}

\begin{thm} \label{thm:informal}
  There is a deterministic one-to-one pushdown transducer that can compress
  some normal sequence.
\end{thm}
A more precise statement is given in Proposition~\ref{pro:formal} where
the pushdown transducer and the normal sequence compressed by it are made
explicit.

\section{Precise statement}

Before giving a more precise statement, we recall a few definitions.  Let
$A$ be a finite alphabet.  Let $A^*$ and $A^ℕ$ be respectively the set of
finite words and the set of (infinite) sequences over~$A$. The positions of
words and sequences are numbered starting at~$1$. To denote the symbol at
position~$i$ of a word (respectively sequence) $w$ we write $w[i]$ and to
denote the substring of $w$ from position~$i$ to~$j$ we write $w[i{:}j]$.
The length of a finite word~$w$ is denoted by~$|w|$.  The empty word is
denoted by~$\emptyword$.  For a word $w = a_1 ⋯ a_n$, let $\widetilde{w}$ be
the \emph{reverse} of~$w$ defined by $\widetilde{w} = a_n ⋯ a_1$.  We write
$\#E$ for the cardinality of a finite set~$E$.  For $w$ and $u$ two words,
let us denote by $|w|_u$ the number of possibly overlapping
\emph{occurrences} of~$u$ in~$w$.  A sequence~$x ∈ A^ℕ$ over alphabet~$A$
is \emph{normal} if
\begin{displaymath}
  \lim_{n→∞}\frac{|x[1{:}n]|_w}{n} = \frac{1}{(\#A)^{|w|}}
\end{displaymath}
holds for each word $w ∈ A^*$.

A \emph{pushdown transducer} is made of input and output alphabets $A$
and~$B$, a stack alphabet~$Z$ containing the starting symbol~$z_0$, a
finite state set~$Q$ containing the initial state~$q_0$ and a finite set of
transitions of the form $p,z \trans{a|v} q,h$ where $p,q∈ Q$, $a ∈ A$, $v ∈
B^*$, $z ∈ Z$ and $h ∈ Z^*$.  The states $p$ and~$q$ are the starting and
ending states of the transition.  The symbol~$a$ and the word~$v$ are its
input and output labels.  The stack symbol~$z$ and the word~$h$ are
respectively the symbol popped from the stack and the word pushed to the
stack.  Note that the transition $p,z \trans{a|v} q,h$ replaces the top
symbol~$z$ by the word~$h$.  If $h$ is empty, it just pops the symbol~$z$.
The transducer is \emph{deterministic} if for each triple $(p,z,a)$, there
exists at most one triple $(q,h,v)$ such that $p,z \trans{a|v} q,h$ is one
of its transitions.  Note that pushdown transducers sometimes include
transitions of the form $p,z \trans{\emptyword|v} q,h$, called
$\emptyword$-transitions, that consume no input symbol.  Such transitions
are not needed for our compressor, but are needed for the decompressor as we
shall see.

A configuration~$C$ of the transducer is a pair $\tuple{q,h}$ where $q ∈ Q$
is its state and $h ∈ Z^*$ is its stack content.  Note that the stack
content is written bottom up: the top symbol is the last symbol of~$h$.
The starting configuration is the pair $\tuple{q_0,z_0}$ where $q_0$ is the
initial state and $z_0$ the starting symbol.

A \emph{run step} is a pair of configuration $\tuple{C,C'}$ denoted
$C \trans{a|v} C'$ such that $C = \tuple{p, wz}$, $C' = \tuple{q,wh}$ for
some word $w ∈ Z^*$ and $p,z \trans{a|v} q,h$ is a transition of the
transducer.  A finite (respectively infinite) \emph{run} is a finite
(respectively infinite) sequence of consecutive run steps
\begin{displaymath}
  C_0 \trans{a_1|v_1} C_1 \trans{a_2|v_2} \cdots \trans{a_n|v_n} C_n.
\end{displaymath}
The input and output labels of the run are respectively $a_1 ⋯ a_n$ and
$v_1 ⋯ v_n$.  Note that a transition $p,z \trans{a|v} q,h$ can be seen as a
run step whose starting stack content is reduced to a single symbol~$z$.
Conversely, each run step is obtained from a transition
$p,z \trans{a|v} q,h$ by adding a stack content~$w$ below the top
symbol~$z$.

Let $A$ be the alphabet $\{0, …, k-1\}$ for some positive integer~$k$ and
let $B$ be the alphabet $A ⊎ \{\extraone, \extratwo\}$ where $\extraone$
and $\extratwo$ are two new symbols not in~$A$.  Now we give the
deterministic pushdown transducer~$𝒯_k$ with input alphabet~$A$ and output
alphabet~$B$.  We first describe it informally and second we give a more
formal description of its transitions.  The transducer $𝒯_k$ proceeds as
follows whenever it reads a symbol~$a ∈ A$ from the input tape.  If the
symbol~$a$ is different from the top symbol of the stack, the symbol~$a$ is
pushed onto the stack and it is also written to the output tape.  If the
symbol~$a$ is equal to the top symbol of the stack, this top symbol is
popped.  Every two symbols consecutively popped from the stack, a
symbol~$\extratwo$ is written to the output tape.  An additional
symbol~$\extraone$ is also written to the output tape if the whole sequence
of consecutive popped symbols is of odd length.  In other words, after a
maximal sequence of $n$ consecutive pops, is written to the output tape
either the word $\extratwo^{n/2}$ if $n$ is even or the word
$\extratwo^{(n-1)/2}\extraone$ if $n$ is odd.  This coding of the
length~$n$ is far from being optimal but it is sufficient to get
compression.  More formally the state set of~$𝒯_k$ is $Q = \{0, 1\}$ and the
initial state is $q_0=0$.  Its
stack alphabet is $A ⊎ \{⊥\}$ and the start symbol~$z_0$ is the new
symbol~$⊥$.  As the symbol~$⊥$ is different from any input symbol, it is
never popped from the stack.  Therefore, the symbol~$⊥$ always remains at
the bottom of the stack and it is used to mark it.  The transitions set~$E$
of~$𝒯_k$ is defined as follows.

\begin{alignat*}{2}
  E = & \{ 0,z \trans{a|a} 0,za : z ≠ a \}
      & \qquad & \text{Pushing $a$ and outputting $a$} \\
      & \{ 0,z \trans{a|\emptyword} 1,\emptyword : z = a\} 
      & \qquad & \text{Popping $z = a$ and outputting $\emptyword$} \\
      & \{ 1,z \trans{a|\extratwo} 0,\emptyword : z = a\} 
      & \qquad & \text{Popping $z = a$ and outputting $\extratwo$} \\
      & \{ 1,z \trans{a|\extraone a} 0,za: z ≠ a \}
      & \qquad & \text{Pushing $a$ and outputting $\extraone a$} 
\intertext{
The function realized by this transducer is one-to-one and the inverse
function can even be computed by the following deterministic pushdown
transducer. This transducer works as follows. Each symbol $a\in A$ is pushed
to the stack and output. When $\extraone$ is read, one symbol from
the stack is popped and output. When $\extratwo$ is read, two symbols from
the stack are popped and output (the topmost first).
}
  E' = & \{ 0,z \trans{a|a} 0,za : z,a\in A \}
      & \qquad & \text{Pushing $a$ and outputting $a$} \\
      & \{ 0,z \trans{\extraone|z} 0,\emptyword : z\in A \} 
      & \qquad & \text{Reading $\extraone$, popping and outputting}\\
&     & \qquad & \text{the top stack symbol} \\
      & \{ 0,z \trans{\extratwo|z} 1,\emptyword : z\in A \} 
      & \qquad & \text{Reading $\extratwo$, popping and outputting}\\
&     & \qquad & \text{the top stack symbol} \\
      & \{ 1,z \trans{\emptyword|z} 0,\emptyword : z\in A \}
      & \qquad & \text{Popping and outputting}\\
&     & \qquad & \text{the top stack symbol again} 
\end{alignat*}

Let $𝒯$ be a pushdown transducer with input alphabet~$A$ and output
alphabet~$B$.  The \emph{compression ratio}~$ρ$ of an infinite run
\begin{displaymath}
  C_0 \trans{a_1|v_1} C_1 \trans{a_2|v_2} C_2 \trans{a_3|v_3} \cdots
\end{displaymath}
is
\begin{displaymath}
  ρ = \liminf_{n→∞}\frac{|v_1 ⋯ v_n|\log\#B}{n\log\#A}
\end{displaymath}
The factors $\log\#A$ and $\log\#B$ take into account the alphabet sizes.
Without them, it would be to easy to compress by taking a larger
alphabet~$B$.  The transducer~$𝒯$ is said to \emph{compress} a sequence~$x$
if it realizes a one-to-one function and if the compression ratio~$ρ$ of
the infinite run of~$𝒯$ on input~$x$ satisfies $ρ < 1$.

The following proposition is a more precise reformulation of 
Theorem~\ref{thm:informal}.
\begin{prop} \label{pro:formal}
  Let $A$ be the alphabet $\{0, …, k-1\}$ for some large enough integer~$k$.
  Let $w_n$ be, for each integer $n ⩾ 1$, the concatenation in lexicographic
  order of all words of length~$n$ over~$A$.  The deterministic pushdown
  transducer~$𝒯_k$ given above compresses the normal sequence
  $x = w_1\widetilde{w}_1w_2\widetilde{w}_2w_3\widetilde{w}_3⋯$.
\end{prop}

Before proving the proposition, we make some comments.  The proof that the
sequence~$x$ is normal is an easy adaptation that the Champernowne sequence
is normal~\cite[Thm~7.7.1]{BecherCarton18}.

The proposition states the result for $k$ large enough.  The proof below
shows that the condition $k ⩾ 7$ is sufficient but numerical experiments
show that $k ⩾ 5$ is actually sufficient.

Some other normal sequences are compressible by the same transducer.  For
each integer $n ⩾ 1$, let $u_1,…,u_{ℓ_n}$ be an enumeration in some order
of all words of length~$n$ over~$A$.  This means that $ℓ_n = (\#A)^n$.  Let
$w_n$ be the word $u_1\widetilde{u}_1u_2\widetilde{u}_2⋯
u_{ℓ_n}\widetilde{u}_{ℓ_n}$ for each integer $n ⩾ 1$.  The sequence $x =
w_1w_2w_3⋯$ is also compressible by the same transducer~$𝒯_k$.  It seems
that this result can be proved using the same techniques.  However, our
numerical experiments suggest that the compression ratio of this latter
sequence is worse than the one given in the proposition.

Our numerical experiments show that the compression ratio converges
to~$3/4$ when the alphabet size~$k$ goes to infinity. It seems that the
same ideas used in the proof of the proposition can achieve this result,
but we preferred simplicity in our presentation.

\section{Proof}

Now we introduce a congruence~$∼$ on~$A^*$ which is used to characterize
stack contents of the pushdown transducer~$𝒯_k$.  Let $→$ be the relation
defined on~$A^*$ as follows.  Two words $w$ and~$w'$ satisfy $w → w'$ if
there are two words $u$ and~$v$ and a symbol $a ∈ A$ such that $w = uaav$
and $w' = uv$.  The word~$w'$ is thus obtained from~$w$ by deleting two
consecutive identical symbols.  A word~$w$ is \emph{irreducible} for~$→$ if
it contains no consecutive occurrences of the same symbol.  Let $\trans{*}$
be the reflexive-transitive closure of the relation~$→$.  Let us recall
that the relation~$→$ is \emph{Noetherian} if there is no infinite
chain $w_0 → w_1 → w_2 → ⋯$ and that it is \emph{confluent} if the
relations $w \trans{*} w_1$ and $w \trans{*} w_2$ imply that there exists
another word~$w'$ such that $w_1 \trans{*} w'$ and $w_2 \trans{*} w'$.

\begin{lem}
  The relation $→$ is Noetherian and confluent.
\end{lem}
\begin{proof}
  Since $w → w'$ implies $|w| > |w'|$, the relation~$→$ is obviously
  Noetherian. Hence, by Newman's lemma, it is sufficient for confluence to
  prove that $→$ is locally confluent. This means that relations $w → w_1$ and
  $w → w_2$ imply that there exists $w'$ such that
  $w_1 \trans{*} w'$ and $w_2 \trans{*} w'$. Suppose that $w → w_1$ and $w → w_2$ where $w_1$
  and~$w_2$ are obtained from~$w$ by deleting respectively the blocks $aa$
  and~$bb$ of two identical symbols. Either the two blocks overlaps and
  $a = b$ or they are disjoint. In the former case, $w$ is equal to
  $u_1aaau_3$ for $u_1,u_3 ∈ A^*$ and in the latter case, $w$ is equal to
  $u_1aau_2bbu_3$ for $u_1,u_2,u_3 ∈ A^*$ if it is assumed, by symmetry, that
  $aa$ occurs before~$bb$. In the former case $w' = w_1 = w_2 = u_1au_3$ and in the
  latter case $w_1 = u_1u_2bbu_3$ and $w_2 = u_1aau_2u_3$ and then $w_1 → w'$
  and $w_2 → w'$ where $w' = u_1u_2u_3$.
\end{proof}

The fact that $→$ is Noetherian and confluent implies that for each
word~$w$, there is a unique irreducible word~$\widehat{w}$ such that $w
\trans{*} \widehat{w}$.  Let us define the equivalence relation~$∼$
on~$A^*$ by $w ∼ w'$ if and only if $\widehat{w} = \widehat{w}'$.  It can
be checked that the relation~$∼$ is the reflexive-symmetric-transitive
closure of~$→$, that is, the relation $(→ ∪ ←)^*$: the equality
$\widehat{w} = \widehat{w}'$ implies the relations $w \trans{*} \widehat{w}
= \widehat{w}' \tranr{*} w'$ and the converse is due to confluence which
allows us to replace each pattern $w_1 \tranr{*} w \trans{*} w_2$ by the
pattern $w_1 \trans{*} w' \tranr{*} w_2$ for some word~$w'$.  The
equivalence relation~$∼$ is actually a congruence: if $u ∼ u'$ and $v ∼
v'$, then $uv ∼ u'v'$. Note that each palindrome of even length, that is,
each word of the form~$w\widetilde{w}$, satisfies $w\widetilde{w} ∼
\emptyword$.  The following lemma is easily proved by induction on the
length of~$w$.

\begin{lem} \label{lem:stack}
  After reading a word~$w$, the stack content of~$𝒯_k$ is $⊥\widehat{w}$
  where $\widehat{w}$ is the unique irreducible word such that $w \trans{*}
  \widehat{w}$.
\end{lem}

Let us recall that the input sequence is
$w_1\widetilde{w}_1w_2\widetilde{w}_2w_3\widetilde{w}_3⋯$.  The lemma just
stated above implies that the stack only contains the bottom symbol~$⊥$
after reading the prefix $w_1\widetilde{w}_1⋯ w_n\widetilde{w}_n$ because
$w_i\widetilde{w}_i ∼ \emptyword$ for each integer $i ⩾ 1$.

Let $P = \{1,…,|w_n\widetilde{w}_n|\}$ be the set of positions of symbols
in $w_n\widetilde{w}_n$.  Each symbol of $w_n\widetilde{w}_n$ is consumed
by either a pushing transition or a popping transition.  In the former
case, the consumed symbol is pushed to the stack.  In the latter case, the
same symbol as the one consumed is popped from the stack.  This dichotomy
induces the partition $P = P_0 ⊎ P_1$ where $P_0$ is the set of positions
of symbols being pushed and $P_1$ is the set of positions of symbols
popping.  Since the stack only contains the bottom symbol~$⊥$ before and
after reading $w_n\widetilde{w}_n$, each pushed symbol is popped later.
Then, the run of~$𝒯_k$ also induces a function $f$ from~$P_0$ to~$P_1$
which maps each position of a pushed symbol to the position of the symbol
that pops it.  This function~$f$ is of course, one-to-one and onto because
each pushed symbol is popped by exactly one symbol.  By definition, the
function~$f$ satisfies that $i < f(i)$ for each $i$ in~$P_0$ and that the
symbols at positions $i$ and $f(i)$ are the same.  The stack policy implies
that if two positions $i$ and~$j$ in~$P_0$ satisfy $i < j$, then $f(i) >
f(j)$.  Let us call an \emph{edge} a pair $(i,f(i))$.  An edge is
\emph{short} if $f(i) - i = 1$ and is \emph{long} if $f(i) - i > 1$.

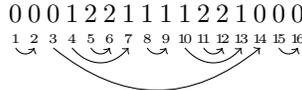
\begin{figure}[htp]
  \begin{center}
    \begin{tikzpicture}[yscale=2.8, xscale=1.8]
    \foreach \x/\xtext in {0/0,0.25/0,0.5/0,0.75/1,1/2,1.25/2,1.5/1,1.75/1}
      \node[anchor=base] at (\x,0) {$\scalebox{1.2}{\xtext}$};
    \foreach \x/\xtext in {0/1,0.25/2,0.5/3,0.75/4,1/5,1.25/6,1.5/7,1.75/8}
      \node[anchor=base] at (\x,-0.2) {$\scriptstyle\xtext$};
    \foreach \x/\xtext in {2.0/1,2.25/1,2.5/2,2.75/2,3/1,3.25/0,3.5/0,3.75/0}
      \node[anchor=base] at (\x,0) {$\scalebox{1.2}{\xtext}$};
    \foreach \x/\xtext in {2.0/9,2.25/1\!0,2.5/1\!1,2.75/1\!2,3/1\!3,3.25/1\!4,3.5/1\!5,3.75/1\!6}
      \node[anchor=base] at (\x,-0.2) {$\scriptstyle{\xtext}$};
    \path[->] (0.00,-0.25) edge[bend left=-50] (0.25,-0.25);
    \path[->] (0.50,-0.25) edge[bend left=-30] (3.25,-0.25);
    \path[->] (0.75,-0.25) edge[bend left=-40] (1.50,-0.25);
    \path[->] (1.00,-0.25) edge[bend left=-50] (1.25,-0.25);
    \path[->] (1.75,-0.25) edge[bend left=-50] (2.00,-0.25);
    \path[->] (2.25,-0.25) edge[bend left=-40] (3.00,-0.25);
    \path[->] (2.50,-0.25) edge[bend left=-50] (2.75,-0.25);
    \path[->] (3.50,-0.25) edge[bend left=-50] (3.75,-0.25);
    \end{tikzpicture}
  \end{center}
  \caption{Example of a function $f$: $f(3) = 14$.}
\end{figure}

Let us call a \emph{block} a maximal set $\{i, i+1, …, j\}$ of consecutive
positions with the same symbol at each position.  Maximal means here that
the set cannot be expanded to the left because either $i = 1$ or symbols at
positions $i-1$ and~$i$ are different and that it cannot be expanded to the
right because either $j = |w_n\widetilde{w}_n|$ or symbols at positions $j$
and~$j+1$ are different.  The following lemma states a link between the
number of long edges and the length of the output of~$𝒯_k$.

\begin{lem} \label{lem:output}
  While reading $w_n\widetilde{w}_n$, the transducer~$𝒯_k$ writes at most
  $|w_n\widetilde{w}_n| - h/6$ symbols where $h$ is the number of blocks of
  length~$1$ in $w_n\widetilde{w}_n$.
\end{lem}
\begin{proof}
  Let $d$ be the difference between the length of $w_n\widetilde{w}_n$ and
  the number of symbols written by~$𝒯_k$ while reading
  $w_n\widetilde{w}_n$.  We have to prove that $d ⩾ h/6$.  Each symbol
  pushed to the stack by~$𝒯_k$ is also written to the output tape.  A
  maximal sequence of $n$ consecutive popping transitions of~$𝒯_k$ writes
  $\extratwo^{n/2}$ if $n$ is even and $\extratwo^{(n-1)/2}\extraone$ if
  $n$ is odd.  This shows that such a maximal sequence of length $n ⩾ 2$
  contributes $⌊ n/2 ⌋ ⩾ n/3$ to~$d$.
  
  Let $N$ be the number of popping transitions belonging to a sequence of
  at least two popping transitions.  From the previous reasoning $d ⩾
  N/3$.
  
  For each block of length~$1$, there is a long edge $(i,f(i))$ such that
  either $i$ or $f(i)$ belongs to the block.  This shows that the number of
  long edges is at least $h/2$.  Due to the nesting of edges, the position
  $f(i)-1$ is also the arrival of another edge.  These two edges contribute
  at least $1$ to~$N$.  This shows that $N ⩾ h/2$ and hence $d ⩾ h/6$.
\end{proof}

\begin{lem} \label{lem:oddblocks}
  For $n ⩾ 3$, the number of blocks of length~$1$ in $w_n\widetilde{w}_n$
  is exactly
  \begin{displaymath}
    \frac{(k-1)^2}{k^2}|w_n\widetilde{w}_n|.
  \end{displaymath}
\end{lem}
\begin{proof}
  Note that $w_n$ starts with $n$ occurrences of the symbol~$0$ and ends
  with $n$ occurrences of the symbol~$k-1$.  It follows that a block of
  length~$1$ in $w_n\widetilde{w}_n$ can occur neither at the beginning nor
  at the end of $w_n$ and~$\widetilde{w}_n$.  The number of blocks of
  length~$1$ in $w_n\widetilde{w}_n$ is twice the number of blocks of
  length~$1$ in~$w_n$.
  
  If each word of length $n$ has exactly $m$ cyclic occurrences in a word $w$,
  then each word~$u$ of length $1\leqslant \ell\leqslant n$ has $mk^{n-\ell}$
  occurrences since $u$ is the prefix of $k^{n-\ell}$ words of length $n$. By
  Theorem~5 in \cite{AlBeFeYu16}, each word of length~$n$ has exactly $n$
  cyclic occurrences in~$w_n$. Applying the previous remark for $\ell=3$
  yields that each word of length~$3$ has exactly $nk^{n-3}$ cyclic
  occurrences in~$w_n$. A block of length~$1$ corresponds to an occurrence of
  a word $abc$ where the symbols $a,b,c ∈ A$ satisfy $a ≠ b$ and $b ≠ c$. Such
  a word $abc$ cannot overlap the border of~$w_n$. It follows that each word
  $abc$ with $a ≠ b$ and $b ≠ c$ has exactly $nk^{n-3}$ occurrences in~$w_n$.
  Since the length of~$w_n$ is $nk^n$ and there are $k(k-1)^2$ such words
  $abc$, the proof is complete.
\end{proof}

\begin{proof}[Proof of Proposition~\ref{pro:formal}]
  Combining Lemmas~\ref{lem:output} and~\ref{lem:oddblocks} yields that,
  for $n ⩾ 3$, the number of symbols written by~$𝒯_k$ while reading the
  word $w_n\widetilde{w}_n$ is at most $(1 -
  (k-1)^2/6k^2)|w_n\widetilde{w}_n|$.  Therefore the transducer~$𝒯_k$ given
  above compresses the normal sequence $x =
  w_1\widetilde{w}_1w_2\widetilde{w}_2w_3\widetilde{w}_3⋯$ as soon as the
  following inequality holds.
  \begin{displaymath}
    \left(1-\frac{(k-1)^2}{6k^2}\right) \frac{\log(k+2)}{\log k} < 1
  \end{displaymath}
  The first term of the left hand side decreases to~$5/6$ and the second term
  decreases to~$1$.  The inequality is satisfied for $k ⩾ 7$ since it boils
  down to $9^{43} < 7^{49}$.
\end{proof}

\paragraph{Acknowledgements.}
We would like to thank the anonymous reviewers for their valuable comments
that improved the quality of the paper.

\bibliographystyle{alphaurl}
\bibliography{pushdown}

\end{document}